\documentclass[reprint, amsmath, amssymb, aps]{revtex4-2}

\usepackage{graphicx}					
\usepackage{bm}							
\usepackage{xcolor}

\makeatletter                           
\@ifclasswith{revtex4-2}{draft}{        
    \usepackage[notref,notcite]{showkeys}
    
}{}
\makeatother

\newcommand{\nn}{\nonumber}
\newcommand{\gl}{\big{(}}
\newcommand{\gr}{\big{)}}
\newcommand{\eV}{\,\text{eV}}
\newcommand{\GeV}{\,\text{GeV}}
\newcommand{\subt}[1]{_{\text{#1}}}
\newcommand{\supt}[1]{^{\text{#1}}}

\newcommand{\vp}{\varphi}
\newcommand{\vpbar}{\bar{\vp}}
\newcommand{\vphat}{\hat{\vp}}
\newcommand{\rhotil}{\tilde{\rho}}
\newcommand{\Nbar}{\bar{N}}
\newcommand{\mtil}{\tilde{m}}
\newcommand{\tin}{t_{\text{in}}}
\newcommand{\ubar}{\bar{u}}
\newcommand{\eps}{\varepsilon}

\begin{document}

\title{Dark energy evolution from quantum gravity}

\author{Christof Wetterich}
\affiliation{Institut f\"ur Theoretische Physik\\
Universit\"at Heidelberg\\
Philosophenweg 16, D-69120 Heidelberg}

\begin{abstract}

If an ultraviolet fixed point renders quantum gravity renormalizable, the
effective potential for a singlet scalar field -- the cosmon -- can be computed
according to the corresponding scaling solution of the renormalization group
equations. We associate the largest intrinsic mass scale generated by the flow
away from the fixed point with the scale of present dark energy density or even
smaller. This results in a highly predictive scenario for the evolution of
dynamical dark energy. It solves the cosmological constant problem dynamically,
and may be called "quantum gravity quintessence". A first setting without
quantum scale symmetry violation in the neutrino sector could explain the
present amount of dark energy, but fails for the constraints on its time
evolution. In contrast, a logarithmic scale symmetry violation in the beyond
standard model sector responsible for the neutrino masses induces a
non-vanishing cosmon-neutrino coupling in the Einstein frame. This yields a
cosmology similar to growing neutrino quintessence, which could be compatible
with present observations. The small number of unknown parameters turns the
scaling solution for quantum gravity into a fundamental explanation of dynamical
dark energy which can be falsified.

\end{abstract}

\maketitle



Already in the first paper on dynamical dark energy or quintessence~\cite{CWQ}
quantum scale symmetry~\cite{CWQS} has played a central role. In quantum gravity
the fate of quantum scale symmetry is closely related to the presence of an
ultraviolet fixed point of the renormalization group equation, which may be
asymptotically safe~\cite{WEI, REU} or asymptotically free~\cite{STE, FRT,
AVBAR, SWY}. For a world living exactly on the fixed point quantum scale
symmetry is typically an exact symmetry. In contrast, for a crossover between
two fixed points, or in case of a flow away from the ultraviolet fixed point due
to a relevant parameter, an intrinsic mass scale yields an explicit breaking of
quantum scale symmetry or dilatation symmetry. This intrinsic mass scale can be
viewed as dimensional transmutation of running couplings, somewhat similar to
the confinement scale in quantum chromodynamics. The largest intrinsic mass
scale sets the overall scale of the model and has no directly observable
meaning. Only dimensionless ratios, as particle masses or field values in units
of this mass scale, are observable. In quantum gravity the largest intrinsic
mass scale is often associated with the Planck mass. There is no need for this,
however, since a dynamical Planck mass may be given by the value of a scalar
field rather than being an intrinsic scale. We propose to associate the largest
intrinsic mass scale with a much smaller scale. It may be of the order of a few
meV -- a scale characteristic for the masses of neutrinos and the present dark
energy density, or even much smaller. This simple assumption leads to a highly
predictive setting for dynamical dark energy~\cite{CWIQ, CWQGSS} that we call
"quantum gravity quintessence".

The basic reason for predictivity is the observation that for momentum scales or
field values much larger than the intrinsic scale the latter becomes negligible.
In this case all properties can be computed from the scaling solution of the
renormalization flow. Scaling solutions are particular solutions of systems of
non-linear differential equations. Their existence imposes severe constraints
which encode properties that may not be recognized by too simple "naturalness
arguments" for the role of quantum fluctuations. The renormalization group
equations or flow equations deal already with the quantum effective action which
includes all effects of quantum fluctuations. The flow equations, and therefore
their particular scaling solutions, are due to quantum effects.

In functional renormalization the dimensionless couplings are not only functions
of a renormalization scale $k$. They are also functions of the cosmon field
$\chi$. For example, the dimensionless effective potential of the cosmon
$u=V(\chi,k)/k^4$ depends, in general, both on $k$ and on $\chi$. For a scaling
solution $u$ is only a function of the dimensionless ratio $\chi/k$. Once this
function is computed, all cosmon self-interactions, quartic, sextic and so on,
are given. Similarly, the coefficient of the curvature scalar $F$ (effective
squared Planck mass) defines a dimensionless ratio $f=F/k^2$. For the scaling
solution $f$ is only a function of $\chi/k$. The same holds for the coefficient
of the kinetic term of the cosmon $K$. If $u$, $f$ and $K$ can be computed, the
cosmological field equations derived by variation of this quantum effective
action lead to typical models of "variable gravity"~\cite{CWVG}.

Already the first computations of models of a singlet scalar field coupled to
the metric ("dilaton quantum gravity") have revealed a simple behavior for large
$\chi/k$~\cite{DQG1, DQG2}. The function $f$ increases quadratically with
$\chi$, $f\sim\chi^2/k^2$, such that $F=\xi\chi^2$. The dimensionless coupling
$\xi$ is the non-minimal coupling of the scalar field to gravity. It is found
not to vanish. On the other hand, for the scaling solution the potential settles
to a constant for large $\chi$, $u(\chi/k)\to u_\infty$, $V=u_\infty k^4$. The
scaling solution describes a crossover from an ultraviolet fixed point for
$\chi/k\to0$ to an infrared fixed point for $\chi/k\to\infty$. For a scaling
solution describing a crossover the intrinsic scale is set by $k$. The intrinsic
scale has a dominant effect for the scalar potential $V\sim k^4$, while for the
coefficient of the curvature scalar it remains subdominant,
$F\sim\xi\chi^2+f_\infty k^2$.

The present cosmological epoch is very close to the infrared fixed point, since
the ratio $\chi/k$ has grown to huge values $\sim10^{30}$ during the long
history of the universe. This has crucial consequences for predictivity. First,
the effects of the metric fluctuations are tiny, being suppressed by powers of
$k^2/\chi^2$. (An exception is the contribution to a constant term in the scalar
effective potential.) Second, for momenta or field values much larger than the
neutrino mass the intrinsic scale $k$ becomes negligible. In this momentum range
one obtains the scale invariant standard model~\cite{CWQ, SHAZEN1, SHAZEN2},
with all mass scales (e.g. Fermi scale, confinement scale) proportional to
$\chi$. This proportionality is a simple consequence of quantum scale symmetry
which becomes exact for $k\to0$. The particles and couplings of the standard
model are well known, such that the renormalization group equations can be
computed without additional unknown parameters. The functional flow equations
coincide with the ones from perturbation theory.

A loophole for predictivity arises from possible beyond standard model (BSM)
physics. Even though the BSM sector may not involve additional light particles,
it manifests itself through the masses of neutrinos. They are due to dimension
five operators which involve a heavy scale where the symmetry B-L is
spontaneously broken. We explore here two cases. For the first, the neutrino
masses are exactly proportional to $\chi$. In this case the model is very
predictive for cosmology. The field equations for the cosmon admit solutions for
which the present amount of dark energy can be obtained. This is already rather
remarkable since no small dimensionless parameter is present, and the tiny ratio
$V/F^2\sim10^{-120}$ arises from the dynamical increase of $\chi/k$ over the
history of the universe. The detailed equation of state is not compatible with
observation, however.

For our second case we consider the possibility that quantum scale symmetry
violation in
the BSM-sector induces an additional dependence of neutrino masses on $\chi/k$.
This dependence will become computable only once a given BSM-sector is assumed.
At the present stage, we parameterize the logarithmic running in the BSM-sector
by a single free parameter, such that the model still remains highly predictive.
The outcome is a model close to growing neutrino quintessence~\cite{CWGN, ABW}
which may well be compatible with present observation.

\subsection*{Cosmon potential from quantum gravity}

Quantum gravity computations of the flowing effective action by use
of functional flow equations typically yield for a range of fields relevant for
our purpose the quantum effective action according to the scaling solution,
\begin{equation}
\label{A2}
\Gamma_k=\int_x\sqrt{g'}\left\{-\frac\xi2\chi^2R'+u(\chi)k^4+\frac12K
\partial^\mu\chi\partial_\mu\chi\right\}\
.
\end{equation}
The central point for predictivity is the computation of $u(\chi)$ according to
the scaling solution of quantum gravity. We are interested in very large values
of $\chi^2/k^2$. In this range the gravitational fluctuations decouple
effectively for the flow equation for $\xi(\chi)$ and $K(\chi)$. Their
contribution is suppressed by $k^2/\chi^2$, reflecting the inverse squared
Planck mass. Both quantities reach constant values which are determined by the
flow for smaller $\chi^2/k^2$ where the gravitational fluctuations still matter,
and fluctuations for unknown particles beyond the standard model could play a
role. Overall, the understanding of the flow of $K$ and $\xi$ has not yet
reached the same reliability as for $u$.

For the computation of observational consequences it is most suitable to work in
the Einstein frame with a fixed (reduced) Planck mass
$M=2.436\cdot10^{18}\,\text{GeV}$. The field equations follow then from a
quantum effective action
\begin{equation}
\label{A1}
\Gamma=\int_x\sqrt{g}\left\{-\frac{M^2}{2}R+U_E(\vp)+\frac12Z\partial^\mu\vp
\partial_\mu\vp\right\}\
.
\end{equation}
Here the ``wave function renormalization'' or ``kinetial'' $Z$ is, in general,
a function of $\vp$. The Weyl transformation of eq.~\eqref{A2} to the Einstein
frame rescales the metric for constant $\xi$ by 
\begin{equation}
\label{W1}
g_{\mu\nu}^{(E)} = \frac{\xi\chi^2}{M^2}g'_{\mu\nu}\ .
\end{equation}
This results in
\begin{equation}
\label{A3}
U_E(\vp)=\frac{u(\chi)k^4M^4}{\xi^2\chi^4}\ ,
\end{equation}
with scalar field $\chi$ and $\vp$ related by
\begin{equation}
\label{A4}
\vp=4M\ln\left(\frac\chi k\right)\ .
\end{equation}
For constant $\xi$ one has
\begin{equation}
\label{A5}
Z(\vp)=\frac1{16}\left(\frac{K(\chi)}{\xi}+6\right)\ .
\end{equation}
We take here a constant $Z$ as one of the parameters of our model. 

If $u(\chi)$ reaches for asymptotically large $\chi$ a constant $u_\infty$ and
$\xi$ is constant, the potential $U_E(\vp)$ decreases exponentially,
\begin{equation}
\label{A6}
U_E=\frac{u_\infty M^4}{\xi^2}\exp\left(-\frac\vp M\right)\ .
\end{equation}
This amounts to a possible dynamical solution of the cosmological constant
problem~\cite{CWQ} for cosmologies where $\vp$ diverges in the infinite future.
At present, $\vp/M$ is large but finite, resulting in very small $U_E$ as needed
for an explanation of the present value of the dark energy density. For the
scaling solution the Weyl transformation to the Einstein frame eliminates the
renormalization scale $k$. For cosmology one can directly employ the field
equations derived from the effective action~\eqref{A1}.

\subsection*{Functional flow of the cosmon potential}

The ``scaling frame''~\eqref{A2} is the one for which a suitable approximation
(truncation) to the exact functional flow equation~\cite{CWFE, REUW, RGREV,
CWGIF, SFE} is evaluated. The corresponding functional flow equation for the
dimensionless potential $u=V(\chi)/k^4$ describes its dependence on the
renormalization scale $k$, which is given by an infrared cutoff such that in a
Euclidean setting only fluctuations with squared momentum $q^2$ larger than
$k^2$ are included,
\begin{equation}
\label{A7}
k\partial_ku=-4u+2\rhotil\partial_{\rhotil}u+4c_U\ .
\end{equation}
This flow equation holds at fixed $\rhotil=\chi^2/(2k^2)$. The term $-4u$
reflects the denominator in the ratio $u=V/k^4$, the term
$-2\rhotil\partial_{\rhotil}$ results from the transition of the flow at fixed
$\chi$ to fixed $\rhotil$, and the last term $c_U$ describes the effect of
fluctuations on the $k$-dependence of $V(\chi)$. One finds~\cite{PRWY, CWEP}
\begin{equation}
\label{A8}
c_U=\frac1{128\pi^2}\gl\Nbar_S(\rhotil)+2\Nbar_V(\rhotil)-2\Nbar_F(\rhotil)
+\Nbar_g(\rhotil)\gr\ ,
\end{equation}
with $\Nbar_S$, $\Nbar_V$, and $\Nbar_F$ the effective numbers of scalars, gauge
bosons and Weyl-fermions, which depend on $\rhotil$ through $\rhotil$-dependent
particle masses. (For computations of the flow of the scalar potential within
various approximations and truncations see refs.~\cite{DP, NP, PV, DELP, EHLY,
EP, LPF}.) The contributions from non-gravitational fluctuations (matter
fluctuations) are the standard result for functional flow
equations~\cite{CWFE,REUW}, which are well tested in a large variety of
applications~\cite{RGREV}. They include one-loop perturbation theory. In a
version of gauge invariant flow equations~\cite{CWGIF} the contribution from
metric fluctuations can be approximated~\cite{CWGFC} by
\begin{equation}
\label{A9}
\Nbar_g=\frac5{1-v}+\frac1{1-v/4}-4\ ,\quad v=\frac{u}{\xi\rhotil}\ .
\end{equation}
For the large values of $\rhotil$ relevant here $v$ is tiny, resulting in
constant $\Nbar_g=2$, corresponding to the two massless graviton degrees of
freedom.

In the flow equation~\eqref{A8} the contributions of bosons are positive, while
the contributions of fermions are negative. This reflects the well known signs
of the fluctuation contributions to the cosmological constant -- the scalar
potential can be seen as a field-dependent cosmological constant. These
different signs will play an important role for the present note since they are
responsible for a negative potential $U_E(\vp)$ in a certain range of $\vp$.

The effective particle numbers $\Nbar$ involve threshold functions which
account for the decoupling of heavy particles once their squared mass $m^2$ is
larger than the cutoff $k^2$. Details of the threshold functions depend on the
precise form of the cutoff. For the Litim cutoff~\cite{LIT} or the simplified
flow equation~\cite{CWSF} one has for the fermions
\begin{equation}
\label{A10}
\Nbar_F=\sum_f\gl1+\mtil_f^2\gr^{-1}\ ,\quad
\mtil_f^2=\frac{m_f^2(\rhotil)}{k^2}\ .
\end{equation}
Here the sum over $f$ runs over all Weyl- (or equivalently Majorana-) fermions
with mass $m_f$. The fermion masses depend on the scalar field $\chi$ and we
write
\begin{equation}
\label{A11}
m_f=h_f\chi\ ,\quad \mtil_f^2(\rhotil)=2h_f^2(\rhotil)\rhotil\ ,
\end{equation}
with the effective Yukawa couplings $h_f$. Exact quantum scale symmetry implies
constant $h_f$. For example, the effective Yukawa coupling $h_e$ for the
electron is of the form $h_e=y_e\vp_0(\chi)/\chi$, with vacuum expectation
value of the Higgs doublet $\vp_0(\chi)$ proportional to $\chi$ and $y_e$ the
Yukawa coupling to the Higgs doublet. The electroweak gauge hierarchy requires
a tiny value of $\vp_0/\chi$, implying a very small effective Yukawa coupling
$h_e$. From the measured ratio of electron mass $m_e$ to Planck mass $M$ one
infers
\begin{equation}
\label{A12}
h_e=\frac{\sqrt{\xi}m_e}{M}\ .
\end{equation}

In order to compute the relevant value of $\mtil^2$ for a given stage of the
cosmological evolution we need the relevant value of $\rhotil$. The units of
$k$ are arbitrary. We choose units which identify the present value of the
potential with the present observed dark energy density, which obtains for a
present Hubble parameter $h=0.7$ and dark energy fraction $\Omega_{h,0}=0.7$ as
\begin{equation}
\label{A12A}
u(t_0)k^4=U_E(t_0)=(2.229\cdot10^{-3}\,\text{eV})^4\ .
\end{equation}
For realistic cosmology with $U_E(t_0)/M^4\approx10^{-120}$ the present value
of $\chi/k$ must be very large according to $M^2=\xi\chi^2(t_0)$,
\begin{equation}
\label{A13}
U_E(t_0)=\frac{u(t_0)
M^4}{\xi^2}\exp\left(-\frac{\vp(t_0)}{M}\right)=\frac{u(t_0)M^4}{4\xi^2}
\rhotil(t_0)^{-2}\
.
\end{equation}
With $u(t_0)$ a few times $(128\pi^2)^{-1}$, see below, this amounts to a very
large present value of $\rhotil$ somewhere around $\rhotil(t_0)\approx10^{58}$.
This large value will be related later to the increase of $\rhotil(t)$ over a
huge time period in Planck units -- it is a consequence of the huge age of our
universe.

We may estimate the present value of the dimensionless neutrino mass
$\mtil_\nu$,
\begin{equation}
\label{A14}
\mtil_\nu^2(t_0)=\sqrt{u(t_0)}\left(\frac{m_\nu}{2.229\cdot10^{-3}\,\text{eV}}
\right)^2\
.
\end{equation}
This leads to the interesting conclusion that the present epoch of cosmology
coincides roughly with the epoch when the neutrino fluctuations decouple from
the flow of the effective potential. It may therefore not be surprising that
close to the present epoch important qualitative changes can occur in the
evolution of dynamical dark energy. On the other hand, in the present epoch
electrons and all other charged fermions have already decoupled from the flow.

For the boson fluctuations only the photon, the cosmon and the graviton
contribute to the flow of the potential in the range of $\rhotil$ relevant for
the present cosmological epoch. Photons are massless, resulting in $\Nbar_V=1$.
For the cosmon fluctuations the mass term is given by the second derivative of
the effective potential with respect to $\chi$,
\begin{equation}
\label{A15}
\Nbar_S=\frac1{1+u'+2\rhotil u''}\ ,
\end{equation}
where primes denote derivatives with respect to $\rhotil$,
$u'=\partial_{\rhotil}u$. This completes the flow generator $c_U$ for the range
of $\rhotil$ relevant for cosmology close to the present epoch. For earlier
epochs $\rhotil(t)$ is smaller than $\rhotil(t_0)$. Electron fluctuations
matter for a range of $\rhotil$ smaller than $h_e^{-2}$. For even much smaller
$\rhotil$ one has to include fluctuation contributions from muons and pions,
and so on.

\subsection*{Scaling solution for the cosmon potential}

The differential equation~\eqref{A7} admits a scaling solution for
$k\partial_ku=0$. It is given by a solution of the non-linear differential
equation
\begin{equation}
\label{A16}
\rhotil\partial_{\rhotil}u=2(u-c_U)\ .
\end{equation}
The solution of this non-linear differential equation has to exist for the whole
range $0\leq\rhotil<\infty$. Combined with a similar equation for $f(\rhotil)$
this imposes severe constraints, such that the only scaling solutions found have
constant $u$ for $\rhotil\to\infty$. With boundary condition
$u(\rhotil\to\infty)=u_\infty$ the scaling solution reads~\cite{CWQGSS}
\begin{equation}
\label{A17}
u=\frac5{128\pi^2}-\frac1{64\pi^2}\sum_ft_u\gl\mtil_f^2\gr\ ,
\end{equation}
with integrated threshold function obeying the differential equation
\begin{equation}
\label{EE19}
\rhotil\partial_{\rhotil} t_u = 2t_u - \frac{2}{1+\mtil_f^2}\,.
\end{equation}
It describes the effective decoupling of fluctuations with $\mtil_f^2\gg1$.
In the range of $\rhotil$ relevant for present cosmology only the neutrinos
contribute effectively in the sum over fermions. For the cosmon contribution we
employ
\begin{align}
u'&=-\frac1{64\pi^2}\sum_f\partial_{\rhotil}t_u\gl\mtil_f^2\gr\\
&=\frac1{32\pi^2\rhotil}\sum_f\left(\frac1{1+\mtil_f^2} -
t_u\gl\mtil_f^2\gr\right)\ ,\nn
\label{A19}
\end{align}
and infer that the scalar mass is tiny in the relevant range of $\rhotil$,
$\Nbar_S=1$. The factor $5$ in eq.~\eqref{A17} simply counts the number of
massless bosonic degrees of freedom. (We have neglected in the flow equation
mixing effects between metric- and scalar fluctuations~\cite{CWMY}. They do not
alter the conclusion $\Nbar_g+\Nbar_S=3$ for the relevant range of large
$\rhotil$. The scaling solution for $u$ should be seen together with a scaling
solution for the dimensionless coefficient of the curvature scalar $f$. The
solution $f=2\xi\rhotil$ for large $\rhotil$ is already incorporated in our
ansatz~\eqref{A2}. In particular, there is no combined scaling solution for $u$
and $f$ which is compatible with $u(\rhotil\to\infty)\sim\rhotil^2$.)

The scaling solution~\eqref{A17} can be translated to the potential in the
Einstein frame~\eqref{A3}
\begin{equation}
\label{A20}
U_E=\frac{u(\vp)M^4}{\xi^2}\exp\left(-\frac\vp
M\right)=U_0\frac{u(\vp)}{u_\infty}\exp\left(-\frac{\vp-\vpbar}{M}\right)\ ,
\end{equation}
with $u_\infty=5/(128\pi^2)\approx0.004$. Here $U_0$ is a free constant that
specifies the definition of $\bar\vp$ and will be chosen as
$U_0=(2.229\cdot10^{-3}\eV)^4$. The constant $\vpbar$ accounts for the dominant part
of $\vp(t_0)$ and absorbs $\xi$,
\begin{equation}
\label{A21}
\frac\vpbar M=\ln\left(\frac{u_\infty M^4}{\xi^2U_0}\right)\ .
\end{equation}
The present value $\vp(t_0)$ is given by
\begin{equation}
\label{A22}
\frac{\vp(t_0)-\vpbar}{M}=\ln\left(\frac{u\gl\vp(t_0)\gr}{u_\infty}\right)
-\ln\left(\frac{U_E(t_0)}{U_0}\right)\ ,
\end{equation}
where a realistic cosmology requires $0<u\gl\vp(t_0)\gr<u_\infty$. With
eqs.~\eqref{A3},~\eqref{A16} the extrema for $U_E$ occur for the scaling
solution at $c_U = 0$, or
\begin{equation}
\label{23A}
\sum_\nu \frac{1}{1+\mtil_\nu^2} = \frac52\,.
\end{equation}

Besides the overall exponential factor the $\vp$-dependence of $U_E(\vp)$ is
governed by
\begin{equation}
\label{A23}
\frac{u(\vp)}{u_\infty}=1-\frac25\sum_ft_u\gl\mtil_f^2(\vp)\gr\ ,
\end{equation}
with ($2\rhotil=\exp(\vp/2M)$)
\begin{equation}
\label{A24}
\mtil_f^2(\vp)=h_f^2(\vp)\exp\left(\frac{\vp}{2M}\right)=c_f(\vp)\exp\left
(\frac{\vp-\vpbar}{2M}\right)\
.
\end{equation}
The Weyl transformation to the Einstein frame also rescales the fermion masses,
which obey now
\begin{equation}
\label{25A}
m_f(\vp) = h_f(\vp)M/\sqrt{\xi}\,.
\end{equation}
The function
\begin{equation}
\label{A25}
c_f(\vp)=\frac{h_f^2(\vp)}{h_f^2\gl\vp(t_0)\gr}\frac{m_f^2(t_0)\sqrt{u_\infty}}
{\sqrt{U_0}}\
,
\end{equation}
becomes a constant if the Yukawa coupling $h_f$ does not depend on $\rhotil$,
\begin{equation}
\label{A26}
c_{f,0}=\left(\frac{m_f(t_0)}{9\cdot10^{-3}\,\text{eV}}\right)^2\ .
\end{equation}
In the Einstein frame the renormalization scale $k$ is no longer present.

We first concentrate on the case of constant $h_f$ and discuss a possible
$\vp$-dependence in the second part. In this case one finds for the integrated
threshold function
\begin{equation}
\label{27A}
t_u(\mtil_f^2) = 1 - 2\mtil_f^2 -
2\mtil_f^4\ln\left(\frac{\mtil_f^2}{1+\mtil_f^2}\right)\,.
\end{equation}
The decoupling for large $\mtil_f^2\gg1$ follows from
\begin{equation}
\label{27B}
t_u\gl\mtil_f^2\gg1\gr = \frac{2}{3\mtil_f^2}\,.
\end{equation}

\begin{figure}[h]%
\centering
\includegraphics[width=\linewidth]{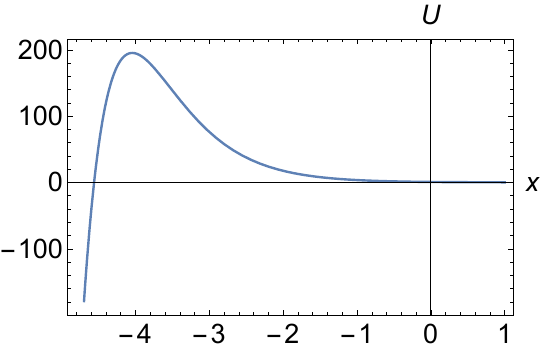}
\caption{Cosmon potential in the Einstein frame for the scaling solution with
hierarchical neutrino masses. We plot $U_E/U_0$ as a function of
$x=(\vp-\bar\vp)/(2M)$.}
\label{fig:1}
\end{figure}

As $\vp$ decreases the dimensionless fermion masses $\mtil_f$ get smaller and
$t_u$ approaches one. For small enough $\vp$ this leads to a negative value of
$u(\vp)$, and therefore to a negative potential $U_E(\vp)$. This negative region
sets in once the largest neutrino mass term $\mtil_\nu^2(\vp)$ is small enough.
For the range of $\vp$ for which $\mtil_e^2\gg1$, $\mtil_\nu^2\ll1$ one has
$\Nbar_f=3$ and therefore $u(\vp)=-1/(128\pi^2)$. The value of $\vphat$ where
the potential switches from negative to positive values depends on the
generation structure of the neutrino masses. For degenerate neutrino masses it
is given by $t_u\gl\mtil_\nu^2(\vphat)\gr=5/6$ or $\mtil_\nu^2\approx1/12$. For
hierarchical neutrino masses, $\mtil_1^2~,\mtil_2^2\ll\mtil_3^2$,
$m_3=0.06\,\text{eV}$, the switch of sign occurs for $t_u\gl\mtil_3^2\gr=1/2$.
For hierarchical neutrino masses we plot the potential $U_E(\vp)$ for the
scaling solution and constant Yukawa couplings in fig.~\ref{fig:1}.
Units of $U_E$ are given by $U_0$, and $x=(\vp-\vpbar)/(2M)=\ln\rhotil-c_\rho$.
The maximum of the potential occurs for values rather close to the present dark
energy density. For degenerate neutrino masses the potential maximum occurs for
smaller $x$ and is higher.

For a given set of neutrino masses the potential $U_E$ for the scaling solution
with constant Yukawa couplings is a parameter free function of the field
variable $(\vp-\vpbar)/M$. It only depends on the masses of neutrinos. In
addition, the dynamics depends on the kinetial $Z$. (We could use a canonical
scalar field $\sigma$. This would lead to a one-parameter family of potentials
$V(\sigma)=U_E\gl\vp(\sigma)\gr$, with $\vp-\vpbar=\sigma/\sqrt{Z}$.) In the
Einstein frame the fermion masses are constant in case of field-independent
Yukawa couplings.

\subsection*{Early cosmology for the scaling potential}

For constant particle masses the evolution of the energy density of particles
(radiation plus matter) follows the standard conservation laws. For a
homogeneous isotropic universe the cosmological field equations read, with $t$
cosmic time, $a(t)$ the scale factor and $H(t)=\partial_t\ln a$ the Hubble
parameter,
\begin{align}
(\partial_t^2+3H\partial_t)\vp&=-\frac1Z\frac{\partial U_E}{\partial\vp}\
,\nn\\
\partial_t\rho_E&=-nH\rho_E\ ,
\label{A27}
\end{align}
with
\begin{equation}
\label{A27A}
H^2=\frac1{3M^2}\left[U_E+\frac Z2(\partial_t\vp)^2+\rho_E\right]\ ,
\end{equation}
and $\rho_E$ the energy density of matter and radiation in the Einstein frame
(without the contribution of the cosmon). One has $n=4$ for radiation
domination, $n=3$ for matter domination, and more generally $n$ is a smooth
function of particle masses and temperature interpolating between the limits.

We are interested in solutions of these field equations for a recent
cosmological epoch for which $\rho_E$ is dominated by non-relativistic matter,
$n=3$. For this purpose we need for some suitable time $\tin$ the initial
values $\vp(\tin)$, $\partial_t\vp(\tin)$ and $\rho_E(\tin)$. We choose
$\vp(\tin)$ in a range where neutrinos are the only fermions contributing to the flow
equation for the cosmon potential. For establishing a range of reasonable values
for $\vp(\tin)$ and $\partial_t\vp(\tin)$ we need to understand the qualitative
features of the cosmological evolution of the scalar field prior to $\tin$. For
realistic cosmologies the contributions of the scalar potential and kinetic
energies will be very small at $\tin$, such that the dominant cosmology near
$\tin$ is matter dominated with $\rho_E(\tin)=3M^2$ $H^2(\tin)=4M^2/(3\tin^2)$.

The outcome of the investigation of early cosmology is rather simple. It
establishes that for a suitable range of $Z$ the initial value $\vp(\tin)$ is
large enough such that $x$ is on the right of the maximum in fig.~\ref{fig:1}.
This amounts to a type of "thawing dark energy"~\cite{CWQ, LIN} for which $\vp$
remains constant for a long period in the radiation and matter dominated epochs.
The gradient of the potential induces a change of $\vp$ only recently. We next
present some details leading to this conclusion.

While we focus in this note on late cosmology, our model actually covers the
whole history of the universe. For small $\rhotil$ the scaling solution for $u$
may involve particles beyond the standard model, as for grand unified theories.
The overall features are rather independent of the precise particle content. An
early inflationary epoch requires positive $U_E$ for a range of small $\chi$.
For the scaling solution of the flow equations the effects of additional bosons
beyond the standard model have to turn $c_U$ positive, c.f. eq.~\eqref{A8}. This
is typically the case for grand unified models, see ref.~\cite{CWQGSS} for a
detailed discussion of the inflationary epoch. It is followed by an epoch of
kination~\cite{CWQ}, for which the energy density of the universe is dominated
by the kinetic energy of the scalar field. Neglecting $\rho_E$ and $U_E$ the
solution of eq.~\eqref{A27} reads
\begin{equation}
\label{A28}
H=\frac1{3t}\ ,\quad \frac\vp
M=\sqrt{\frac{2}{3Z}}\ln\left(\frac{t}{t\subt{kin}}\right)
+\frac{\vp_{\text{kin}}}{M}\
,
\end{equation}
where the kination epoch starts at $t\subt{kin}$ with a value
$\vp(t\subt{kin})=\vp_{\text{kin}}$. For this epoch one has
\begin{equation}
\label{A29}
H^{-2}\exp\left(-\frac{\vp}{M}\right)\sim t^{2-\sqrt{\frac{2}{3Z}}}\ .
\end{equation}
This factor governs the relative importance of the potential $U_E$. In
particular, for $Z<1/6$ the potential becomes more and more negligible during
the epoch of kination. In contrast, during kination the energy density in
radiation or matter decreases slower than the scalar kinetic energy density,
\begin{equation}
\label{A29A}
\rho_E\sim t^{-\frac n3}\ ,
\end{equation}
and will finally overwhelm the latter. If the potential remains negligible the
epoch of kination ends once $\rho_E$ starts to dominate. For the radiation and
matter dominated epochs the kinetic and potential energy of the scalar field
make only a small contribution to $H^2$, such that the overall cosmology
follows the standard picture. As long as $U_E$ can be neglected the approximate
solution becomes
\begin{equation}
\label{A30}
H=\frac{2}{nt}\ ,\quad \dot{\vp}=c_nt^{-\frac6n}\ ,\quad
\vp=\frac{c_nn}{n-6}t^{\frac{n-6}{n}}+\vp_r\ ,
\end{equation}
and the relative importance of the scalar kinetic energy decreases
\begin{equation}
\label{A31}
H^{-2}\dot{\vp}^2\sim t^{\frac{2(n-6)}{n}}\ .
\end{equation}
The scalar field almost stops its evolution, such that the ratio of potential
to kinetic energy of the scalar field increases.

As long as the gradient of the potential can be neglected the overall picture
of the evolution of the scalar field is rather simple. During the kination
epoch $\vp$ increases logarithmically until it almost settles at $\vp_r$ at
some time $\bar t_r$ after the onset of radiation domination. If $\vp_r$ is
sufficiently below the value $\vp_{\text{max}}$ where $U_E$ has its maximum,
the value of the scalar field will start to decrease once the term $\partial
U_E/\partial\vp>0$ becomes important. On the other hand, for
$\vp_r>\vp_{\text{max}}$ the scalar field increases and the potential $U_E$
remains positive for all $t>\bar t_r$. A positive present value of $U_E$ is
needed for any realistic cosmology. A viable solution is therefore
$\vp_r>\vp_{\text{max}}$.

The value $\vp_r$ is determined by the value of the scalar field at the end of
inflation and the duration of the kination epoch. During kination the scalar
field changes by
\begin{align}
\frac{\Delta\vp}{M}&=\sqrt{\frac{2}{3Z}}\ln\left(\frac{t_r}{t_{\text{kin}}}
\right)=\sqrt{\frac{2}{3Z}}\ln\left(\frac{H_{\text{kin}}}{H_r}\right)\nn\\
&=\sqrt{\frac{8}{3Z}}\ln\left[\left(\frac{\rho_{\text{kin}}}{\rho_r
}\right)^{1/4}\right]\
.
\label{N1a}
\end{align}
This change has to be large enough such that $\vp_r>\vp\subt{max}$. On the other
hand, for very large $\vp_r/M$ the cosmon contribution to the energy density
remains very small even at the present epoch. In this case the model cannot
account for dark energy. These constraints limit the allowed range for $Z$.

The measured amplitude of the primordial fluctuations yields information about
the value $\vp$ at the time of decoupling of the primordial density
fluctuations,
\begin{equation}
\label{N2a}
3.56\cdot10^{-8}r=\frac{u(\vp_d)}{\xi^2}\exp\left(-\frac{\vp_d}{M}\right)\ ,
\end{equation}
with $r\lesssim0.05$ the tensor to scalar ratio. Parameterizing the kinetic
energy at the beginning of kination,
$\rho\subt{kin}=(Z/2)(\partial_t\vp(\tin))^2$, in terms of the potential energy
at decoupling,
\begin{equation}
\label{N3a}
\rho_{\text{kin}}=\frac{\ubar_{\text{kin}}}{\xi^2}\exp\left(-\frac{\vp_d}{M}
\right)M^4\
,
\end{equation}
one obtains
\begin{align}
\frac{\vp_r}{M}&=\left(1-\frac1{\sqrt{6Z}}\right)\ln10^9+\sqrt{\frac{8}{3Z}}
\ln\gl2.4\cdot10^{18}\gr\nn\\
&\quad-\sqrt{\frac{8}{3Z}}\ln\left(\frac{\rho_r^{1/4}}{1\,\text{GeV}}\right)
+\tilde\Delta_{\text{trans}}\
.
\label{N4a}
\end{align}
Here $\tilde\Delta_{\text{trans}}$ accounts for details at the end of inflation
and reads in our parameterization, for $r=1/36$,
\begin{equation}
\label{N5a}
\tilde\Delta_{\text{trans}}=\ln\left(\frac{u_d}{\xi^2}\right)+\frac1{\sqrt{6Z}}
\ln\left(\frac{\ubar_{\text{kin}}}{\xi^2}\right)\
.
\end{equation}
Up to smaller quantitative details the value $\vp_r$ where the scalar field
settles after the beginning of radiation domination depends on the two
parameters $Z$ and $\rho_r^{1/4}/1\,\text{GeV}$.

We may roughly estimate the value of $\vp_r$ needed for a realistic cosmology
by associating $U_E(\vp_r)$ with the present dark energy density
$U_0=\gl2.229\cdot10^{-3}\,\text{eV}\gr^4$,
\begin{equation}
\label{N6a}
\frac{u_\infty}{\xi^2}\exp\left(-\frac{\vp_r}{M}\right)M^4=U_0\ .
\end{equation}
This yields, up to shifts of the order of a few,
\begin{equation}
\label{N7a}
\frac{\vp_r}{M}=277=21+\frac{61}{\sqrt{Z}}-\sqrt{\frac{8}{3Z}}\ln\left(
\frac{\vp_r^{1/4}}{1\GeV}\right)\
.
\end{equation}
For $\rho_r^{1/4}=10^6\GeV$ ($1\GeV$) one finds typical values $Z=0.022$
($0.057$). We emphasize that no particular fine tuning of $Z$ is needed for
realistic cosmology. A change of a factor $10^4$ in $U_E(\rho_r)$ corresponds
to an additive change of $\vp/M$ by $9.2$ or a relative change $\Delta Z/Z$ of
around $0.4\sqrt{Z}$. The values of $Z$ found obey $Z<1/6$, such that the
potential remains indeed negligible during kination.

\subsection*{Dynamical dark energy}

We next consider the evolution of the cosmon field in late cosmology, when the
gradient of the potential starts to play a role. It is convenient to employ
variables
\begin{align}
y&=\ln a\ ,\ \ \partial_t=H\partial_y\ ,\ \
x=\frac1{2M}(\vp-\vpbar)=\ln\rhotil-c_{\rho}\ ,\nn\\
\Omega_V&=\frac{U_E}{3M^2H^2}=\frac{U_0}{3M^2H^2}\frac{u(x)}{u_\infty}\exp(-2x)\
,\nn\\
\Omega_K&=\frac{Z(\partial_t\vp)^2}{6M^2H^2}=\frac{2Z}{3}\gl\partial_yx\gr^2\
,\quad \Omega_E=\frac{\rho_E}{3M^2H^2}\ ,\nn\\
\Omega_E&+\Omega_V+\Omega_K=1\ ,\quad \Omega_h=\Omega_V+\Omega_K\ ,
\label{A32*}
\end{align}
for which the scalar field equation reads
\begin{equation}
\label{A32}
\left[\partial_y^2+\gl3+\partial_y\ln
H\gr\partial_y\right]x=\frac3{2Z}\left(1-\frac12\partial_x\ln u\right)\Omega_V\
.
\end{equation}
Taking a $y$-derivative of the equation for $H^2$ yields
\begin{align}
\partial_y\ln H&=-\left(\frac n2\Omega_E+2Z\gl\partial_yx\gr^2\right)\nn\\
&=-\frac12\big[n(1-\Omega_V)+(6-n)\Omega_K\big]\ .
\label{A33}
\end{align}
We therefore have two field equations~\eqref{A32},~\eqref{A33} for the
functions $x(y)$ and
\begin{equation}
\label{A34}
g(y)=\ln\left(\frac{H}{H_0}\right)\ ,
\end{equation}
with
\begin{equation}
\label{A35}
\Omega_V=\frac{U_0}{3M^2H_0^2}\frac{u(x)}{u_\infty}\exp\big\{-2(g+x)\big\}\ .
\end{equation}
Note that $U_0$ and $H_0$ are not parameters of the model. They only define the
convention which fixes the additive constants in the definition of the variables
$x$ and $g$. We choose $H_0^2$ such that $U_0/(3M^2H_0^2)=0.7$.

For a realistic cosmology we have to require that the cosmon energy density
$\rho_h = U_E + (Z/2)(\partial_t\vp)^2$ remains positive for all time. The
monotonic decrease of $\rho_h$ implies that negative $\rho_h$ cannot turn
positive again. The evolution of $\Omega_h$ obeys
\begin{align}
\partial_y\Omega_h=-\partial_y\Omega_E\,&=\Omega_E\gl
n\Omega_V-(6-n)\Omega_K\gr\nn\\
&=\Omega_E\gl n(1-\Omega_E)-6\Omega_K\gr\ .
\label{C1}
\end{align}
The equation of state of the cosmon dark energy is defined by
\begin{equation}
\label{C2}
w_h=\frac{\Omega_K-\Omega_V}{\Omega_K+\Omega_V}=-1+\frac{2\Omega_K}{\Omega_h}\ .
\end{equation}
The condition for realistic cosmology requires for all $t$ the relation
$\Omega_K+\Omega_V>0$ and therefore $w_h>-1$. For all epochs with $U_E<0$,
$\Omega_V<0$ one has $w_h>1$. We can write eq.~\eqref{C1} in terms of $w_h$ as
\begin{equation}
\label{C3}
\partial_y\Omega_h=(1-\Omega_h)\Omega_h\gl n-3(1+w_h)\gr\ .
\end{equation}
The condition for realistic cosmology is equivalent to finite $w_h$ for all
time.

For the scaling solution one has $\partial_x\ln u=2-2c_U/u$, or
\begin{equation}
\label{A36}
\big[\partial_y^2+(3+\partial_yg)\partial_y\big]x=\frac{U_0c_U(x)}
{2ZM^2H_0^2u_\infty}\exp\big\{-2(g+x)\big\}\ .
\end{equation}
Particular solutions with a static field occur for $x(y)=x\subt{max}$, with
$c_U(x\subt{max})=0$ corresponding to the maximum of $U_E(\vp)$. For the Hubble
parameter one has in this case the standard solution with a cosmological
constant
\begin{equation}
\label{A37}
\lambda\subt{max}=\frac{U_0u(x\subt{max})}{u_\infty}\exp(-2x\subt{max})\ ,
\end{equation}
namely
\begin{equation}
\label{A38}
\bar
g(y)=\frac12\ln\left(\frac{\lambda\subt{max}+\rho_{E,0}\exp(-ny)}{3M^2H_0^2}
\right)\ ,
\end{equation}
which corresponds to $3M^2H^2=\rho_E+\lambda\subt{max}$. This particular
"cosmological constant solution" requires tuning of parameters or initial
conditions such that $\vp_r=\vp\subt{max}$. It is not compatible with
observation since the value $U_E(\vp\subt{max})$ is larger than observed.

\begin{figure}[h]
\centering
\includegraphics[width=\linewidth]{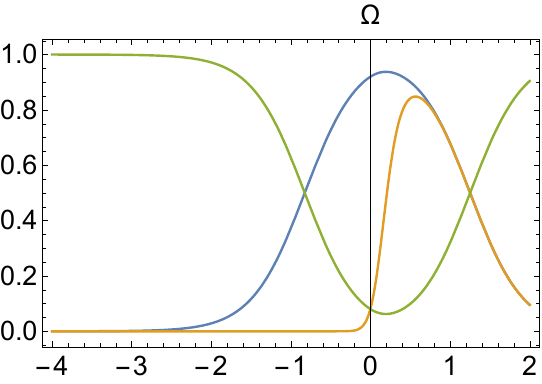}
\caption{Evolution of energy fractions as functions of $y=ln(a)$, for $Z=0.022$.
Matter dominates early cosmology, with $\Omega_E$ (green curve) close to one.
For $y>-2$ the dark energy fraction $\Omega_h$ (blue curve) starts to increase.
At present ($y=0$) it is dominated by the potential, as visible from the small
$\Omega_K$ (orange curve).}
\label{fig:2}
\end{figure}

\begin{figure}[h]
\centering
\vspace{1em}
\includegraphics[width=\linewidth]{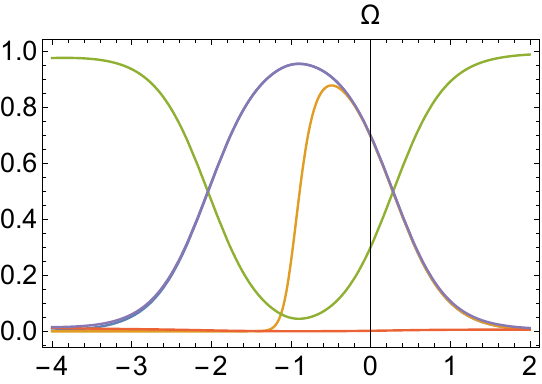}
\caption{Evolution of energy fractions as function of $y=\ln a$ for $Z=0.022$.
The color coding is the same as for Fig.~\ref{fig:2}. The present dark energy
fraction is $\Omega_h=0.7$, with equation of state $w_h(0)$ close to one. The
dynamics is similar to Fig.~\ref{fig:2}, but shifted in $y$.}
\label{fig:3}
\end{figure}

We have solved the system of evolution equations numerically choosing
hierarchical neutrino masses $m_\nu=(0.0005,\,0.008,\,0.058)\eV$ and $Z=0.022$.
The maximum of $U_E$ occurs for $x\subt{max}\approx-4.042199$. We take at
$y\subt{in}=\ln\left(\frac1{500}\right)$ the initial values
$\partial_yx(y\subt{in})=0$, $g(y\subt{in})=9.28777$. If we start with
$x(y\subt{in})$ slightly above $x\subt{max}$ the scalar field has changed only
little at $y=0$. Thus dark energy is dominated by the cosmon potential and one
finds for a typical example the present values for dark energy
$\Omega_h(0)=0.922$ and $w_h(0)=-0.983$. The value of $\Omega_h(0)$ is too high
as compared to the observed value $\Omega_h(0)\approx0.7$. This is due to the
fact that $U_E(x\subt{max})$ is larger than $U_0$. Increasing further
$x(y\subt{in})$, the present scalar kinetic energy and $w_h$ increase, resulting
for example in $w_h(0)=-0.822$, $\Omega_h(0)=0.92$. We plot for this case the
evolution of $\Omega_h$, $\Omega_K$ and $\Omega_E$ in fig.~\ref{fig:2}. In the
future for $y>0$ a scaling solution~\cite{CWQ, CWCMCC} with constant
$\Omega_h=3Z$ and $w_h=0$ will be reached.
As one further increases $x(y\subt{in})$ slightly, dark energy becomes important
earlier, and also the decrease of $\Omega_h$ and increase of $\Omega_K$ begin at
smaller $y$. One can tune $x(y\subt{in})$ in order to obtain a given
$\Omega_h(0)$. For $x(y\subt{in})=-4.04219658$ the evolution is shown for
$\Omega_h$, $\Omega_K$ and $\Omega_E$ in fig.~\ref{fig:3}. The present dark
energy fraction $\Omega_h(0)=0.7$ would be consistent with observation. At $y=0$
the dark energy (blue curve) is dominated by the cosmon kinetic energy (orange
curve), resulting in an equation of state $w_h(0)\approx1$. This is not
compatible with observation. The present value of the Hubble parameter obtains
from $g(0)=0.0022$ very close to $H_0$.

We draw three main conclusions from this investigation. First, for the scaling
solution of quantum gravity with constant neutrino masses in the Einstein frame
the overall features of cosmology look qualitatively similar to a typical
universe with thawing dark energy, provided $Z$ is in a suitable range. Second,
the precise values of $\Omega_h(0)$ and $w_h(0)$ conflict with observation.
Third, with our given assumptions the scaling solution of quantum gravity is
predictive. Besides the neutrino masses and $Z$ there is no free parameter
available which could influence the outcome. For larger neutrino masses and
somewhat different $Z$ the incompatibility with observation does not change.

There are several ways how the cosmology corresponding to the scaling solution
of quantum gravity can be modified. First for the scaling solution the effective
neutrino Yukawa couplings $h_\nu(\rhotil)$ may depend on $\rhotil$
non-trivially. We will see that this can lead to growing neutrino
quintessence~\cite{CWGN, ABW}.

Second, the solution of the flow equation may deviate from the scaling
solution. This implies the presence of relevant parameters and intrinsic mass
scales associated to them. The largest intrinsic mass scale may be denoted by
$\bar k$. As a consequence, the scaling solution holds to a good approximation
for $k\gg\bar k$, while substantial deviations from the scaling solution occur
for $k\leq\bar k$. The maximal intrinsic mass scale is a free parameter which
sets the overall scale. We take it in the vicinity of $2\cdot10^{-3}\eV$. The
effective squared Planck mass will then be dominated by $M^2\approx\xi\chi^2$,
with only a tiny correction $\sim\bar k^2$. For the scaling potential $U=uk^4$
one expects more substantial modifications, for example by an additional
constant $\bar k^4$. A discussion of this possibility is postponed to future
work.

\subsection*{Scale symmetry violation in the neutrino sector}

Let us consider the possibility that the effective neutrino Yukawa coupling
$h_\nu(\rhotil)$ shows a non-trivial dependence on $\rhotil$. The masses of the
left-handed neutrinos of the standard model of particle physics arise from
non-renormalizable couplings, which are sensitive to some ``beyond standard
model'' (BSM) sector. A non-trivial $\rhotil$-dependence of the scaling
solution in the BSM-sector, beyond the proportionality of mass scales to
$\chi$ according to quantum scale symmetry, can result in non-trivial
$h_\nu(\rhotil)$. Thus $\partial_{\rhotil}h_\nu(\rhotil)\neq0$ indicates a
violation of quantum scale symmetry for the scaling solution in the BSM-sector.

If the characteristic $\chi$-dependent mass scales in the BSM-sector are much
larger than the Fermi scale (expectation value of the Higgs doublet
$\vp_0\sim\chi$), the neutrino masses are small as compared to the electron mass
due to some ``seesaw mechanism'' or ``cascade mechanism''. They are suppressed
either by the mass of a ``right-handed'' or ``sterile'' neutrino~\cite{MIN, YAN,
GRS}, or by the mass of a scalar triplet~\cite{MaW, LSW}. If one of these masses
is not exactly proportional to $\chi$, or dimensionless couplings are not
independent of $\chi$, the quantum scale symmetry in the BSM-sector is violated.
We make here the simplified assumption that this scale symmetry violation is
common to all three neutrino masses and parameterize
\begin{equation}
\label{N1}
m_\nu=b_\nu\frac{\vp_0^2(\chi)}{m_{\text{B}-\text{L}}(\chi)}
=\frac{b_\nu\eps\chi^2}{g_{\text{B}-\text{L}}(\chi)\chi}\ ,
\end{equation}
with constant $b_\nu$. Here $\vp_0$ is the doublet expectation value,
$\vp_0^2=\eps\chi^2$, $\eps\ll1$, and
$m_{\text{B}-\text{L}}(\chi)=g_{\text{B}-\text{L}}(\chi)\chi$ is the effective
heavy mass scale associated to $\text{B}-\text{L}$-violating effects in the
BSM-sector. The effective coupling $g_{\text{B}-\text{L}}(\chi)$ incorporates
the details of the mass-generation for neutrinos. For the scaling solution one
has
\begin{equation}
\label{N2}
h_\nu(\rhotil)=\frac{b_\nu\eps}{g_{\text{B}-\text{L}}(\rhotil)}\ ,
\end{equation}
such that a non-trivial $\rhotil$-dependence arises from the non-trivial
$\rhotil$-dependence of $g_{\text{B}-\text{L}}(\rhotil)$.

Typically, $g_{\text{B}-\text{L}}$ is some dimensionless combination of
couplings, as the Yukawa coupling of the right-handed neutrinos, or quartic
scalar couplings which determine the mass of the heavy triplet, a cubic coupling
between the triplet and two doublets, or ratios of various expectation values.
For a scaling solution which has not yet settled at some quantum scale invariant
limit one could expect some logarithmic dependence of $g_{\text{B}-\text{L}}$ on
$\chi/k$, which we parameterize in the range of interest for $\rhotil$ by
\begin{equation}
\label{N3}
g_{\text{B}-\text{L}}(\rhotil)=\bar
g_{\text{B}-\text{L}}-c_{\text{B}-\text{L}}\ln\left(\frac\chi k\right)\ .
\end{equation}
(This parameterization is not supposed to be valid for $\rhotil\to0$ or in a
range where $g_{\text{B}-\text{L}}(\rhotil)$ would vanish.) Translating to $\vp$
yields
\begin{equation}
\label{N4}
h_\nu(\vp)=\frac{4b_\nu\eps M}{4\bar
g_{\text{B}-\text{L}}M-c_{\text{B}-\text{L}}\vp}\ .
\end{equation}
We consider positive $c_{\text{B}-\text{L}}$ and $g_{\text{B}-\text{L}}$ such
that the neutrino masses in the Einstein frame increase with increasing $\vp$.

\subsection*{Cosmon-neutrino coupling}

The $\vp$-dependence of the neutrino masses in the Einstein frame induces an
effective coupling between the cosmon and neutrinos~\cite{CWGN, CWCMCC},
\begin{equation}
\label{N5}
\beta=-\frac{\partial\ln
m_\nu}{\partial\vp}M=-\frac{M}{\vp_c-\vp}=-\frac{1}{2(x_c-x)}\ ,
\end{equation}
with
\begin{equation}
\label{N6}
\frac{\vp_c}{M}=\frac{4\bar g_{\text{B}-\text{L}}}{c_{\text{B}-\text{L}}}\
,\quad x_c=\frac{\vp_c-\bar\vp}{2M}\ .
\end{equation}
For a slow running, $c_{\text{B}-\text{L}}\ll\bar g_{\text{B}-\text{L}}$, the
unknown parameter $\vp_c/M$ may be in the range of the present value of $\vp/M$,
i.e. $\vp_c$ close to $\bar\vp$. In the limit $c_{\text{B}-\text{L}}\to0$,
$x_c\to\infty$ one has $\beta\to0$ and recovers the case of the
field-independent Yukawa coupling discussed previously. The cosmon-neutrino
coupling has been employed in several models of dynamical dark energy with
mass-varying neutrinos~\cite{GZW, FNW, BFLZ, BB, BRO, AZ, BJA, IK}.

The individual neutrino masses obey
\begin{equation}
\label{N6A}
m_\nu(\vp)=\frac{\mu_\nu(\vp_c-\vp_0)}{\vp_c-\vp}=\frac{\mu_\nu(x_c-x_0)}
{x_c-x}\ .
\end{equation}
If we take for $\vp_0$ the present value of $\vp$, and correspondingly for
$x_0$, the free parameters $\mu_\nu$ correspond to the present neutrino masses.
In principle, one can choose $x_0$ arbitrarily. The free parameters in the
neutrino sector are then $x_c$ and the three constant masses $\mu_\nu$, which
together with $x_0$ define the neutrino masses for a given value of $\vp$. In
other words, $x_c$ is the only additional free parameter for this scenario.
With $h_\nu(\vp)=m_\nu(\vp)\sqrt{\xi}/M$ one inserts for the scaling
potential~\eqref{A17}
\begin{equation}
\label{69A}
\tilde m_\nu^2(x)=\sqrt{\frac{u_\infty\mu_\nu^4}{U_0}}\left(
\frac{x_c-x_0}{x_c-x} \right)^2\exp(x)\ .
\end{equation}

A non-vanishing cosmon-neutrino coupling $\beta$ has several effects on the
cosmology of the scaling solution. First, the evolution equation of the cosmon
contains an additional term proportional to the trace of the energy-momentum
tensor of the neutrinos
\begin{equation}
\label{N7}
Z\gl\partial_t^2+3H\partial_t\gr\vp=-\frac{\partial
U_E}{\partial\vp}+\frac{\beta}{M}(\rho_\nu-3p_\nu)\ .
\end{equation}
Second, the energy-momentum tensor of the neutrinos is not conserved, according
to~\cite{CWTVC, CWGN},
\begin{equation}
\label{N8}
\partial_t\rho_\nu+3H(\rho_\nu+p_\nu)=-\frac{\beta}{M}(\rho_\nu-3p_\nu)
\partial_t\vp\ .
\end{equation}
Energy is exchanged between the neutrino- and cosmon-sector and only the
combined energy momentum tensor for the scalar field and neutrinos is
covariantly conserved,
\begin{equation}
\label{N8A}
\partial_t(\rho_\nu+\rho_h)=-3H\gl\rho_\nu+p_\nu+Z(\partial_t\vp)^2\gr\ .
\end{equation}
Third, a variable neutrino mass influences the time when neutrinos get
non-relativistic, i.e. when the neutrino pressure $p_\nu$ differs substantially
from $\rho_\nu/3$. For negative $\beta$ and increasing neutrino masses
neutrinos become non-relativistic later as compared to the case of constant
neutrino masses. Only for non-relativistic neutrinos the modifications in
eqs.~\eqref{N7},~\eqref{N8} matter.

\begin{figure}[h]
\centering
\vspace{1em}
\includegraphics[width=\linewidth]{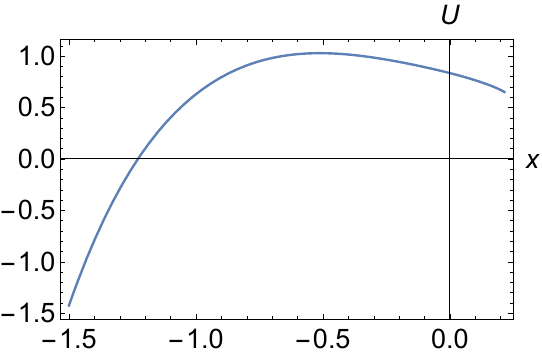}
\caption{Cosmon potential for running neutrino Yukawa coupling. We plot
$U_E/U_0$ as function of $x=(\vp-\bar\vp)/(2M)$ for the parameters $x_c=0.216$,
$x_0=0.08$, $\mu_1=0.002\eV$, $\mu_2=0.01\eV$, $\mu_3=0.24\eV$.}
\label{fig:5}
\end{figure}

Fourth, the scaling solution for the cosmon
potential is modified since $\tilde m_f^2(\vp)$ involves an additional factor
\begin{align}
\label{N9}
\frac{h_\nu^2(\vp)}{h_\nu^2(\vp_0)}&=\left(
\frac{4\bar g_{\text{B}-\text{L}}M-c_{\text{B}-\text{L}}\vp_0}
{4\bar g_{\text{B}-\text{L}}M-c_{\text{B}-\text{L}}\vp(t)}\right)^2\nn\\
&=\left(\frac{\vp_c-\vp_0}{\vp_c-\vp(t)}\right)^2=\left(
\frac{x_c-x_0}{x_c-x}\right)^2\ .
\end{align}
The integrated threshold functions obey now the relation
\begin{equation}
\label{65AA}
\partial_xt_u^{(\nu)} = 2t_u^{(\nu)} - \frac{2}{1+a_\nu(x_c-x)^{-2}e^x}\,,
\end{equation}
with
\begin{equation}
\label{65A}
a_\nu = -\sqrt{\frac{u_\infty\mu_\nu^4}{U_0}}(x_c - x_0)^2\,.
\end{equation}
It has the limiting behavior (with constant $b_-$)
\begin{align}
\label{65B}
t_u^{(\nu)}(x\to-\infty) = 1 - b_-e^{2x}\,,\nn\\
t_u^{(\nu)}(x\to x_c) = \frac{2e^{-x_c}}{3a_\nu}(x_c - x)^3\,.
\end{align}
We solve eq.~\eqref{65AA} numerically. For $x_c=0.216$, $x_0=0.08$,
$\mu_1=0.002\eV$, $\mu_2=0.01\eV$, $\mu_3=0.24\eV$, $\bar\mu_\nu=0.024\eV$ we
plot the effective cosmon potential in
fig.~\ref{fig:5}.
As compared to Fig.~\ref{fig:1} the potential maximum is shifted to larger $x$
and its height is lower. For larger $\mu_\nu$ the potential maximum occurs for
smaller $x$ and its height increases.

Fifth, the additional cosmon-mediated attractive force between neutrinos
accelerates the growth of fluctuations in the neutrino-sector. In our
approximation all these effects are governed by a single new parameter $x_c$,
which determines the function $\beta(x)$ according to eq.~\eqref{N5}.

\subsection*{Cosmology for growing neutrino quintessence}

For cosmology we have to take the effects of the neutrino fluid into account,
with
\begin{equation}
\label{N11}
\Omega_\nu=\frac{\rho_\nu}{3M^2H^2}\ ,\quad
\Omega_m+\Omega_V+\Omega_K+\Omega_\nu+\Omega_\gamma=1\ .
\end{equation}
Here $\Omega_m$ and $\Omega_\gamma$ are the fractions of matter and photons,
corresponding to the energy densities with the usual scaling behavior
$\rho_m\sim a^{-3}$, $\rho_\gamma \sim a^{-4}$.
For the evolution of $\rho_\nu$ or $\Omega_\nu$ we need the effective equation
of state in the neutrino-sector $w_\nu=p_\nu/\rho_\nu$. In addition to
eq.~\eqref{N8} or~\eqref{N8A} we will use
\begin{equation}
\label{N12}
\rho_\nu-3p_\nu=\bar m_\nu n_\nu\ ,
\end{equation}
with total number density of neutrinos $n_\nu \sim a^{-3}$, and average neutrino
mass $\bar m_\nu$,
\begin{equation}
\label{N13}
\bar m_\nu=\frac13\sum_\nu m_\nu\ .
\end{equation}
In early cosmology the neutrino masses are negligible,
$w_\nu=1/3$, and
$\rho_{\nu}\sim T_\nu^4$ is given by the effective neutrino temperature $T_\nu$.
For the present epoch of cosmology $p_\nu$ is of the same order as the energy
density of photons and therefore negligible, $w_\nu\approx0$.

For the present cosmological epoch we may combine the cosmon and neutrino energy
density into a common effective dark energy density $\rho_d$,
\begin{equation}
\label{N16}
\rho_d=\rho_\nu+\rho_h=\rho_\nu+U_E+\frac Z2\gl\partial_t\vp\gr^2\ .
\end{equation}
Neglecting $p_\nu$ one has for late cosmology, with
$\bar\mu_\nu=\frac13\sum_\nu\mu_\nu$,
\begin{align}
\label{N17}
\partial_t\rho_d&=-3H\left[\frac{\bar\mu_\nu
n_\nu(\vp_c-\vp_0)}{\vp_c-\vp}+Z\gl\partial_t\vp\gr^2\right]\nn\\
&=-3(1+w_d)H\rho_d\ .
\end{align}
The equation of state for this combined effective dark energy $w_d$ obeys for
non-relativistic neutrinos
\begin{align}
\label{N18}
w_d&=\frac{T-U_E}{T+U_E+\rho_\nu}\ ,\quad T=\frac Z2\gl\partial_t\vp\gr^2\
,\nn\\
\rho_\nu&=\frac{\vp_c-\vp_0}{\vp_c-\vp}\bar\mu_\nu n_\nu\ .
\end{align}
It is restricted by $w_d>-1$, coming close to $(-1)$ if $T$ and $\rho_\nu$ are
small as compared to $U_E$. This combined effective equation of state can be
compared to the one for the scalar field~\eqref{C2}.

For a numerical solution we employ the field equation
\begin{align}
\label{N19}
&\partial_y^2x+(3+\partial_yg)\partial_yx\nn\\
&=\frac{3}{2Z}\left[\left(1-\frac12\partial_x\ln
u\right)\Omega_V-\frac{\Omega_\nu(1-3w_\nu)}{2(x_c-x)}\right]\ ,
\end{align}
where we insert
\begin{equation}
\label{N19AA}
\partial_yg=-\frac32\gl1+w_\nu\Omega_\nu+\Omega_K-\Omega_V\gr-\frac12
\Omega_\gamma\ .
\end{equation}
Here $\Omega_V$ and $\Omega_K$ are given by eqs.~\eqref{A35} and~\eqref{A32*},
respectively. For the scaling solution we employ
\begin{equation}
\label{73A}
\left(1-\frac12\partial_x\ln u\right)\Omega_V = 0.7\frac{c_U}{u_\infty}
\exp\gl-2(g+x)\gr\,.
\end{equation}

The factor $1-3w_\nu$ suppresses the effects of the
cosmon-neutrino coupling as long as the neutrinos are relativistic. Using
$n_\nu\sim a^{-3}$ we employ
\begin{align}
\label{N19A}
\Omega_\nu(1-3w_\nu)&=\frac{\bar m_\nu n_\nu}{3M^2H^2}\\
&=\frac{\bar\mu_\nu B}{3M^2H_0^2}\exp\big\{-(3y+2g)\big\}
\frac{x_c-x_0}{x_c-x}\,,\nn
\end{align}
with
\begin{equation}
\label{N19D}
B=n_\nu(y\subt{in})\exp(3y\subt{in})=\frac{4c_\nu}{11}T_{\gamma,0}^3\ ,
\end{equation}
where $T_{\gamma,0}=2.348\cdot10^{-4}\eV$ is the present photon temperature and
$c_\nu\approx0.548$, see below. The evolution of the neutrino fraction obeys
\begin{equation}
\label{83A}
\partial_y\Omega_\nu=-\Omega_\nu\left[4+2\partial_yg - \left(1 +
\frac{\partial_y x}{x_c-x}\right)\gl1-3w_\nu\gr\right]\ .
\end{equation}
In turn, we can combine eq.~\eqref{N19A} with $\Omega_\nu$ in order to extract
the neutrino equation of state $w_\nu$ in eq.~\eqref{N19AA},
\begin{equation}
\label{88A}
w_\nu\Omega_\nu=\frac13\Omega_\nu - \frac13\gl1-3w_\nu\gr \Omega_\nu\ .
\end{equation}
For the photon fraction we employ
\begin{equation}
\label{88B}
\Omega_\gamma=\frac{\pi^2T_{\gamma,0}^4}{45M^2H_0^2}\exp \gl(-(4y+2g)\gr\ .
\end{equation}
At this point all quantities for the system of cosmic
equations~\eqref{N19},~\eqref{N19AA},~\eqref{83A} are well defined. This system
is solved numerically.

\subsection*{Initial conditions}

We need to determine the initial conditions, which we choose here at
$a\subt{in}=1/3000$, $y\subt{in}=-8.0$ near matter-radiation equality. The
initial neutrino number density $n_\nu(y\subt{in})$ is given by
($\zeta(3)=1.202$)
\begin{equation}
\label{N19B}
n_\nu(y\subt{in})=c_\nu T_\nu^3(y\subt{in})\ ,\quad
c_\nu=\frac{9\zeta(3)}{2\pi^2}=0.548\ .
\end{equation}
The neutrino temperature $T_\nu$ is related to the photon temperature
$T_\gamma$
\begin{equation}
\label{N19C}
T_\nu=\left(\frac{4}{11}\right)^{\frac13}T_\gamma=\left(\frac{4}{11}\right)
^{\frac13}T_{\gamma,0}\exp(-y)\ .
\end{equation}
This yields eq.~\eqref{N19D} and implies for the evolution
equation~\eqref{N19} of the scalar field
\begin{equation}
\label{N19E}
\Omega_\nu(1-3w_\nu)=D\frac{\bar\mu_\nu(x_c-x_0)}{1\eV}
\frac{\exp\gl-(3y+2g)\gr}{(x_c-x)}\ ,
\end{equation}
with
\begin{equation}
\label{N19F}
D=\frac{4c_\nu T_{\gamma,0}^3(1\eV)}{33M^2H_0^2}=0.0657\ .
\end{equation}

For the initial value of $g(y\subt{in})$ we need the contribution of photons
and neutrinos to the energy density at $y\subt{in}$, as determined for an early
epoch where neutrino masses are negligible by
\begin{equation}
\label{N20}
\Omega_r(a)=\Omega_\gamma(a)+\Omega_\nu(a)=\frac{a\subt{eq}}{a+a\subt{eq}}\ ,
\end{equation}
with matter-radiation equality at $a\subt{eq}=1/3390$. For
$y\subt{in}=\ln(1/3000)=-8.0$ this yields the total radiation fraction
$\Omega_r(y\subt{in})=0.47$. The initial value for $g$ obeys
\begin{align}
\label{N21}
g(y\subt{in})=&\,-\frac{3y\subt{in}}{2}+\frac12\ln\gl1-\Omega_{d,0}\gr\nn\\
&-\frac12\ln\gl1-\Omega_r(y\subt{in})-\Omega_h(y\subt{in})\gr\ ,
\end{align}
Here $\Omega_{d,0}$ is the present total dark energy fraction which sums the
cosmon potential and kinetic energy, as well as the energy density of neutrinos.
For $\Omega_{d,0}=0.7$ this amounts to $g(y\subt{in})=11.708$. The value of
$g(y\subt{in})$ has to be adapted to match the outcome of
$\Omega_{d,0}$. This tuning can be avoided by a more direct formula in
eq.~\eqref{90C} below.

For the initial value of $\Omega_\nu$ we use that the neutrino masses are
negligible in early cosmology such that
\begin{equation}
\label{N23}
\frac{\Omega_\nu}{\Omega_\gamma}=c_{\nu\gamma}=\frac{21}{8}
\left(\frac{4}{11}\right)^{\frac43}\ ,\quad \Omega_\nu=\frac{c_{\nu\gamma}}
{1+c_{\nu\gamma}}\Omega_r\ ,
\end{equation}
and
\begin{equation}
\label{N24}
\Omega_\nu(y)=\frac{c_{\nu\gamma}a\subt{eq}}{(1+c_{\nu\gamma})
(e^y+a\subt{eq})}\ .
\end{equation}
For $y\subt{in}=-8$ one finds the initial value
$\Omega_\nu(y\subt{in})=0.19$.

\subsection*{Evolution of energy densities}

As an alternative numerical setting we follow directly the evolution of
$\rho_m$, $\rho_\gamma$ and $\rho_\nu$ in addition to the evolution of the
scalar field given by eq.~\eqref{N19}. This yields the evolution of the critical
density $3M^2H^2 = \rho_m + \rho_\gamma + \rho_\nu + \rho_h$ and therefore
$g(y)$. Since in contrast to the energy fractions $\Omega_j$ the densities
$\rho_j$ vary by orders of magnitude the coincidence of the numerical results
for both systems of differential equations can be used for an estimate of the
numerical error which turns out to be very small.

The evolution equation~\eqref{N19} for $g$ implies that $\rho_m$, as defined by
$\Omega_m=1-(\Omega_h+\Omega_\nu+\Omega_\gamma)$, scales $\sim a^{-3}$. For
numerical robustness we can implement directly this relation by
\begin{equation}
\label{90A}
\rho_m(y)=\rho_{m,0}\exp(-3y)\ ,
\end{equation}
with $\rho_{m,0}$ determined from the scale factor at radiation-matter equality
as
\begin{equation}
\label{90B}
\rho_{m,0}=(1+c_{\nu\gamma})\rho_{\gamma,0}a\subt{eq}^{-1}\ ,\quad
\rho_{\gamma,0}=\frac{\pi^2}{15}T_{\gamma,0}^4 .
\end{equation}
For the epoch when the neutrino masses are negligible the ratio
$\rho_\nu/\rho_\gamma$ is fixed. For our model $\rho_h$ is negligible around
matter-radiation equality or at last scattering. The Hubble parameter is then
determined for this epoch.

For the physics around last scattering relevant for the observed CMB-anistropies
the two quantities $T_{\gamma,0}$ and $a\subt{eq}$ determine $\rho_m(y)$,
$\rho_\gamma(y)$, $\rho_\nu(y)$ and $H(y)$. With a given baryon ratio
$\Omega_b/\Omega_m$ also the electron energy density $\rho_e(y)$ is known.
Combining the measured value of $T_{\gamma,0}$ with the value of $a\subt{eq}$
from Planck-data ensures that for our choice of parameters all physics around
last scattering is compatible with constraints from CMB-data. This extends up to
the time when neutrino masses begin to matter. One may choose parameters such
that the distance to the last scattering surface is the one measured by the
Planck collaboration. For the CMB observations the difference between our model
and an $\Lambda$CDM model fitting the Planck data concerns then only the physics
in the late epoch when dark energy starts to matter.

For neglected neutrino masses one can infer an expression for $g(y)$,
\begin{align}
\label{90C}
g(y)=\frac12\bigg[&\ln\big(\rho_{m,0}e^{-3y} + \rho_{\gamma,0}(1+c_{\nu\gamma})
e^{-4y}+\rho_h\big)\nn\\
-&\ln\big(3M^2H_0^2\big)\bigg]\ .
\end{align}
We actually use this expression for the determination of $g(y\subt{in})$, which
yields results very close to eq.~\eqref{N21}.

For a numerical solution of the system of differential
equations we treat $x(y\subt{in})$ as a free parameter, set by the evolution
of the scalar field prior to $y\subt{in}$. We find that the evolution is rather
independent of the initial condition for $\partial_yx(y\subt{in})$. After a
rather short initial period $\partial_yx$ settles at a scaling behavior (see
below). We take $(\partial_yx)(y\subt{in})=0$ as second initial condition. The
initial conditions for $g(y\subt{in})$ and $\Omega_\nu(y\subt{in})$ are fixed by
eqs.~\eqref{90C}~\eqref{N24}.

\subsection*{Example for small neutrino masses}

The detailed behavior of the solution of the cosmological field equations for
our model of growing quintessence depends on the parameters in a sensitive way.
We present here an example for small hierarchical neutrino masses. The
parameters are chosen as
\begin{align}
\label{89A}
Z&=0.022\,,\quad x_c=0.216\,,\quad x_0=0.08\,,\quad \bar\mu =
0.024\eV\,,\nn\\
\mu_1&=0.002\eV\,,\quad \mu_2=0.01\eV\,,\quad \mu_3=0.06\eV\,,\nn\\
y\subt{in}&=-8.0\,,\quad x(y\subt{in})=-0.15\,,\quad
\partial_yx(y\subt{in})=0\,.
\end{align}
For dark energy this results in the present energy fractions
\begin{equation}
\label{89B}
\Omega_d(0) = 0.695\,,\quad \Omega_h(0) = 0.694\,,\quad \Omega_\nu(0) = 0.001\,,
\end{equation}
and equation of state
\begin{equation}
\label{89C}
w_d(0) = -0.770\,,\quad w_h(0) = -0.771\,.
\end{equation}
For the Hubble parameters one finds
\begin{equation}
\label{89D}
h(0) = 0.683\,,\quad g(0) = -0.025\,,
\end{equation}
and the distance to the last scattering surface in units of the one measured by
the PLANCK-collaboration is given by
\begin{equation}
\label{89E}
\frac{r_d}{r_d\supt{(Planck)}} = 0.997\,.
\end{equation}
For the neutrino masses one has
\begin{equation}
\label{89F}
x(0) = -0.0013\,,\quad \bar m_\nu(0) = 0.015\eV\,.
\end{equation}

These values are, however, only a snapshot of fast oscillations as a function of
redshift. For example, one has at redshift $z=1$ an equation of state very close
to $-1$, $w_d(z=1)=-0.995$, and much smaller neutrino masses $\bar
m_\nu(z=1)=0.004\eV$. We describe in the following these oscillations in detail.

\begin{figure}[h]
\centering
\vspace{1em}
\includegraphics[width=\linewidth]{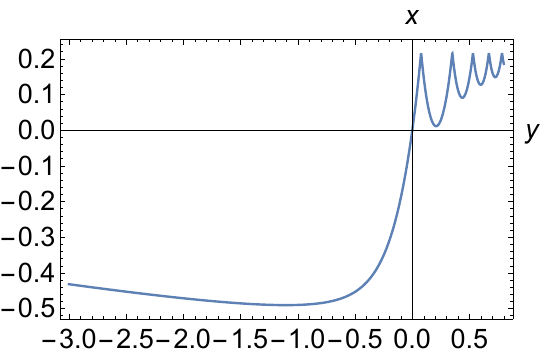}
\caption{Evolution of the dimensionless scalar field $x=(\vp-\bar\vp)/(2M)$ as
function of $y=\ln a$. Parameters are given by eq.~\eqref{89A}, i.e. $Z=0.022$,
$x_c=0.216$, with neutrino masses determined by $\bar\mu_\nu=0.024\eV$,
$x_0=0.08$. For $x(y\subt{in})=-0.15$ at $y\subt{in}=-\ln(3000)$ the cosmon
neutrino coupling is small initially and induces a slow decrease of $x$. As the
universe expands the relative importance of the potential gradient increases,
reversing the evolution of $x$ towards a rapid increase near $y=-0.5$.
Subsequently, the scalar field follows an oscillatory approach towards $x_c$.}
\label{fig:6}
\end{figure}

In fig.~\ref{fig:6} we display the evolution of the dimensionless scalar field
$x$ as a function of
$y=\ln a$.
The neutrino-induced force $\sim\beta$ first drives $x$ slowly towards smaller
values. The values of $x$ shown in the figure correspond to the tail of the
scalar potential at the right of the maximum in fig.~\ref{fig:5}. At some moment
the gradient of the potential overtakes and $x$ is pushed to larger values
again. The competition between the neutrino-induced and gradient force, which
have opposite sign, results in an oscillatory behavior. For time increasing to
the future ($y>0$) the scalar field essentially approaches a constant. The
scalar potential $U_E(x)$ therefore becomes almost constant, which can be
identified with an effective cosmological constant.

The decrease of $x$ before the potential becomes important can be understood by
solving its evolution equation neglecting the contribution from the potential
\begin{equation}
\label{EV1}
\partial_y^2x + (3+\partial_yg)\partial_yx = -A(x_c-x)^{-2}
\exp\gl-(3y+2g)\gr\,,
\end{equation}
with
\begin{equation}
\label{EV2}
A = \frac{3D}{4Z}\frac{\bar\mu_\nu(x_c-x_0)}{1\eV}\,.
\end{equation}
For the matter dominated epoch one has $\partial_yg=-3/2$ and, with
$\Omega_{m,0}\approx0.3$,
\begin{equation}
\label{EV3}
g \approx -\frac32y + \frac12\ln\Omega_{m,0}\,,
\end{equation}
resulting in
\begin{equation}
\label{EV4}
\partial_y^2x + \frac32\partial_yx = -\tilde A(x_c-x)^{-2}\,,\quad \tilde A =
A\Omega_{m,0}\,.
\end{equation}
This is the damped motion of a particle in a potential
\begin{equation}
\label{EV5}
\tilde V(x) = \frac{\tilde A}{x_c-x}\,.
\end{equation}
The ``energy'' decreases according to
\begin{equation}
\label{EV6}
E(y) = \frac12(\partial_yx)^2 + \tilde V(x)\,,\quad \partial_yE =
-\frac32(\partial_yx)^2\,.
\end{equation}
For regions where $\partial_y^2x$ can be neglected the approximate solution
reads
\begin{equation}
\label{EV7}
x(y) = x_c - \big[(x_c-x_m)^3 - 2\tilde Ay_m + 2\tilde Ay\big]^{1/3}\,,
\end{equation}
with $x_m = x(y_m)$ and $y_m$ denoting the onset of matter domination.

For the radiation dominated epoch one has $g = -2y + \text{const}$, such that
the additional factor $\exp(y)$ in eq.~\eqref{EV1} suppresses the driving force
for large negative $y$. A numerical solution of eq.~\eqref{EV1} shows a smooth
transition from constant $x$ to the solution~\eqref{EV7}. As a consequence, the
value of $x$ reached after the onset of radiation domination will be lowered by
an amount $\Delta x$ due to the evolution in the matter dominated epoch. If
$\Delta x$ is too large, the scalar field evolves to values $x<x\subt{max}$ on
the left of the maximum of the potential. In this case it will quickly decrease
further as soon as the potential becomes important. No realistic cosmology is
found in this case. It is possible that the requirement that $x$ stays larger
than $x\subt{max}$ leads to restrictions on the neutrino masses for which a
realistic cosmology can be obtained. We leave this question for further
investigation.

For the present example the oscillations set in only in the future. For other
parameters they start already in the past, such that the present cosmology
undergoes rapid changes in certain quantities. This is the reason why in our
plots we also show the evolution in the future, $y>0$. This gives a flavor of
what can happen for other parameter choices in the present and recent past.

\begin{figure}[h]
\centering
\vspace{1em}
\includegraphics[width=\linewidth]{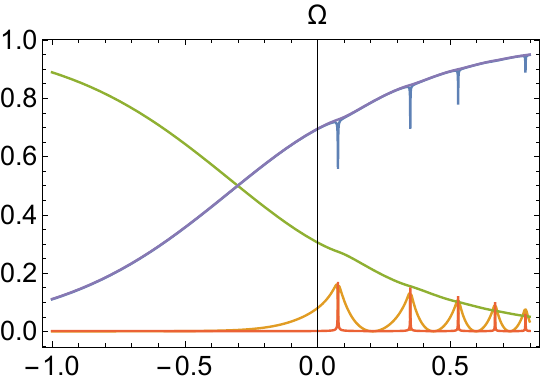}
\caption{Evolution of energy fractions for $Z=0.022$, $x_c=0.216$,
$x(y\subt{in})=-0.15$, $y\subt{in}=-8$. All parameters are the same as
for Fig.~\ref{fig:6}. For the range $y\approx-2$ the universe is dominated by
matter, with $\Omega_m$ (green curve) close to one. For $y\approx-1$ the cosmon
potential starts to matter, leading to an increase of $\Omega_h$ (blue curve).
In this epoch the scalar kinetic energy $\Omega_K$ (orange curve) and the
neutrino energy $\Omega_\nu$ (red curve) are small, such that the combined dark
energy fraction $\Omega_d$ (violet curve) almost coincides with $\Omega_h$.
Subsequently, the rapid increase of $x$ leads to an increase of $\Omega_K$.
Whenever $x$ comes close to $x_c$ the neutrino fraction $\Omega_\nu$ shows
spikes due to the enhanced neutrino masses. These spikes in $\Omega_\nu$ are
accompanied by drops in $\Omega_h$, while $\Omega_d$ shows a relatively smooth
increase, with present value $\Omega_d(0)=0.695$.}
\label{fig:7}
\end{figure}

The evolution of the scalar field is reflected in the evolution of the different
energy densities or corresponding energy fractions $\Omega_m$, $\Omega_d$,
$\Omega_h$, $\Omega_K$ and $\Omega_\nu$ shown in fig.~\ref{fig:7}.
In early cosmology matter and radiation dominate. For the range $y\approx-1$
shown in the figure the universe is matter dominated with $\Omega_m$ (green)
close to one. At this time the cosmon potential starts to play a role, inducing
an increase of $\Omega_V$. As long as the field value changes slowly the kinetic
energy fraction $\Omega_K$ (orange) remains small, such that
$\Omega_h\approx\Omega_V$. Also the neutrino energy fraction $\Omega_\nu$ (red)
is small, such that the curves for $\Omega_h$ (blue) and $\Omega_d$ (violet)
almost coincide. Subsequently, the potential gradient induces a more
rapid increase of $x$. This is the moment when dark energy "thaws", as visible
in the increase of the kinetic energy fraction $\Omega_K$ (orange). The thawing
is stopped, however, once $x$ approaches $x_c$. The strong cosmon-neutrino
coupling $\beta$ counteracts the further increase of $x$ and reverses the sign
of $\partial_yx$. With a decreasing neutrino force the scalar kinetic energy
fraction decreases (sharp drop in $\Omega_K$, orange). The cosmon field
decreases again, until the potential gradient takes over. The oscillations of
$x$ are visible in the oscillations of $\Omega_h$ (blue). The sharp drops in
$\Omega_h$ are partly compensated by sharp peaks in $\Omega_\nu$ which occur
whenever $x$ is close to $x_c$. While $\Omega_h$ and $\Omega_\nu$ oscillate
strongly due to the oscillating scalar field $x$, the combined dark energy
fraction $\Omega_d$ is a more smooth function. For the parameters
chosen $\Omega_d$ reaches a value $\Omega_d(0)=0.695$. In summary, the thawing is
not smooth as for $\beta=0$. The increase of $\Omega_d$, and the corresponding
decrease of $\Omega_m$, show structures which clearly distinguish the evolution
from a cosmological constant.
 
\begin{figure}[h]
\centering
\vspace{1em}
\includegraphics[width=\linewidth]{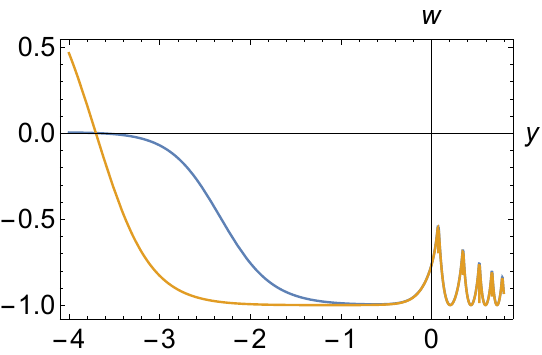}
\caption{Equation of state of scalar field dark energy (orange) and combined
neutrino + scalar dark energy (blue) as function of $y=\ln a$. Parameters are
the same as for Figs.~\ref{fig:6}.}
\label{fig:8}
\end{figure}

In fig.~\ref{fig:8} we show the equation of state for the scalar field energy
density $w_h$ (orange), as well as the effective equation of state $w_d$ for the
combined neutrino-cosmon fluid (blue), as a function of $y$.
\begin{figure}[h]
\centering
\vspace{1em}
\includegraphics[width=\linewidth]{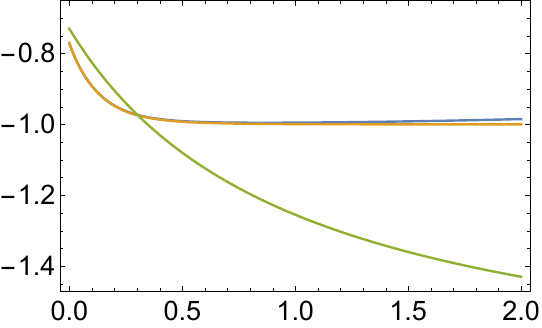}
\caption{Equation of state $w$ as function of redshift $z$. We display $w_h$ for
the scalar field (orange) and the effective equation of state $w_d$ (blue). This
is compared with a $w_0 - w_a$-parameterization~\cite{CHPO} (green) with values
$w_0 = -0.73$, $w_a = -1.05$ from a fit to DESI data~\cite{DESI}. One observes
rough agreement for values $z<0.5$ where dark energy is most important. Values
$w<-1$ are not reached in our model.}
\label{fig:7a}
\end{figure}
For the scalar field $w_h$ starts in early cosmology with positive values
corresponding to the dominance of the kinetic energy. Once the potential becomes
important $w_h$ turns negative, being close to $w_h=-1$ for $y>-3$.
Subsequently, the oscillations of $\Omega_K$ and $\Omega_V$ are reflected in an
oscillating equation of state. For our example the first oscillation peak occurs
in the near future. The equation of state differs substantially from $w_d=-1$
only for redshift $z<0.4$. We compare in Fig.~\ref{fig:7a} the redshift
dependence of our example with a Chevallier-Polarski~\cite{CHPO}
parameterization $w=w_0+w_a(1-a)$ with values $w_0=-0.73$, $w_a=-1.05$,
corresponding to a fit for the DESI data~\cite{DESI}. The behavior for
$z\leq0.4$ is similar, but our model cannot describe a phantom regime with
$w_d<-1$.

\begin{figure}[h]
\centering
\vspace{1em}
\includegraphics[width=\linewidth]{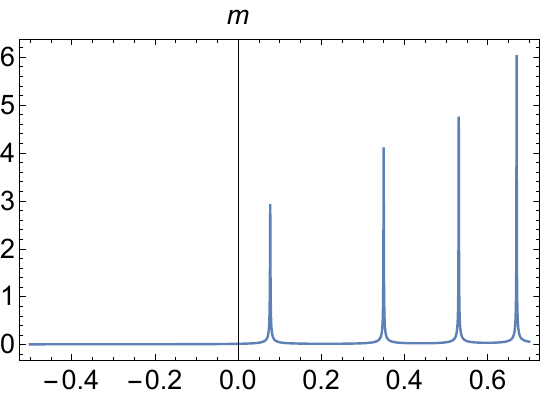}
\caption{Evolution of average neutrino mass $\bar m_\nu$ in units of $\eV$ as
function of $y=\ln a$. The precise location of the peaks depends on the
parameters and initial conditions, given here by eq.~\eqref{89A}.}
\label{fig:9}
\end{figure}

In Fig.~\ref{fig:9} we display the average neutrino mass. Once $x$ becomes
close to $x_c$ (in our example in the future)
the evolution of the neutrino masses is strongly oscillating, with rather sharp
peaks at the turning points for the evolution of the scalar field near $x_c$.
The present average neutrino mass for our particular set of parameters and
initial conditions is $\bar m_\nu=0.015\eV$. The precise value is highly
sensitive to the details in view of the strong oscillations. During structure
formation the neutrino masses have been much smaller.

\begin{figure}[h]
\centering
\vspace{1em}
\includegraphics[width=\linewidth]{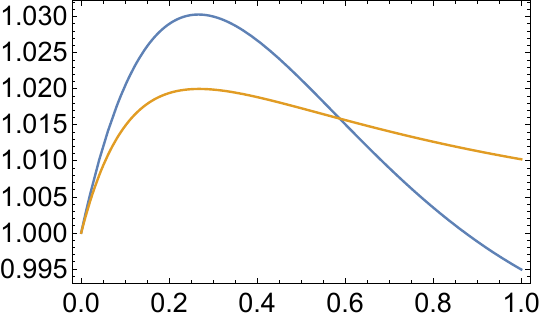}
\caption{Redshift dependence of the Hubble parameter. We plot the Hubble
parameter in units of the Hubble parameter for the $\Lambda$CDM model with the
same value of the present dark energy fraction $\Omega_d(0)$. The orange curve
shows our example with parameters given in eq.~\eqref{89A}, and the blue curve
indicates a Chevallier-Polarski parameterization of $w$ with $w_0=-0.73$,
$w_a=-1.05$.}
\label{fig:11}
\end{figure}

Despite the oscillatory behavior the evolution of the Hubble parameter is rather
smooth. For our parameters and initial conditions the present value is found
$h(0)=0.683$. For other settings $h(0)$ may come out larger than $0.7$. Still,
the detailed evolution of the Hubble parameter shows interesting features. In
Fig.~\ref{fig:11} we show the Hubble parameter for our example divided by the
one for a $\Lambda$CDM model with the same value of the present dark energy
fraction $\Omega_d(0)$. We compare it with a similar ratio for a
Chevallier-Polarski parameterization of $w(a)$ with $w_0=-0.73$, $w_a=-1.05$.
Both show the enhancement between redshifts $z=0$ and $z=0.8$ observed by the
DESI-collaboration~\cite{DESI}.

For other values of the neutrino masses or other values of the parameters $x_c$
and $x(t\subt{in})$ the fast oscillations often occur already in the recent past
for $a<1$, $y<0$. From the overall oscillating picture it is evident that the
detailed value of observables today depends in this case sensitively on the
value of $x_c$ and the initial value $x(\tin)$. In the plane $(x_c,x(\tin))$ one
finds a line for which the present combined dark energy fraction $\Omega_d$
amounts to $\Omega_d = 0.7$. The present distribution of this combined dark
energy on the cosmon potential, cosmon kinetic energy and neutrino energy
density is parameter-dependent. This is seen by the fast oscillations between
these components, and reflected in the oscillating equation of state.
Furthermore, the oscillations depend strongly on the assumed neutrino masses as
encoded in $\bar\mu$ and $x_0$. For larger $\bar\mu$ the oscillations get
slower. The present fraction of neutrino energy density and kinetic cosmon
energy become larger and one starts to see more pronounced oscillations in
$\Omega_h$ and even $\Omega_d$.

A detailed search in parameter space will be necessary in order to find out if
there exists a parameter region for which the model is compatible with all
present observations, possibly overcoming the tensions in the cosmological
constant model. This investigation is not the purpose of the present note.
Furthermore, we should mention that in growing neutrino quintessence the
neutrino fluctuations grow non-linear in a recent cosmological epoch. The
neutrinos form very large lumps, which may render the cosmic neutrino background
observable by large scale inhomogeneities in the gravitational potential. Rather
large backreaction effects are possible. Also the locally observed Hubble
parameter may deviate from its cosmological average, with possible implications
for the Hubble tension. For small neutrino masses these effects are suppressed
by the small neutrino fraction $\Omega_\nu$. Nevertheless, they could play a
certain role for a detailed quantitative analysis. We refer to refs~\cite{MPRW,
WPMW, PWAW, BPAW, AWW, ABF, CPW} for a detailed discussion of these issues.

\subsection*{Conclusions}

In this note we have investigated the consequences of the scaling solution for
quantum gravity for the evolution of dynamical dark energy. Our main assumption
is that the largest intrinsic mass scale produced by the flow of dimensionless
couplings away from the scaling solution is of the order of a few $\text{meV}$
or smaller, rather than the Planck scale as often assumed. This leads to models
of variable gravity and a scale symmetric standard model. The scaling solution
of quantum gravity predicts a very light scalar field -- the cosmon -- as the
pseudo Goldstone boson of spontaneously broken quantum scale symmetry. For
suitable parameters, i.e. an appropriate range for $Z$, the evolution of the
cosmon field induces dynamical dark energy. This is a striking prediction. It
will be crucial to find out if the required value of $Z$ is compatible with the
scaling solution for the scalar kinetic term.

The second central outcome states that the scaling solution of quantum gravity
is highly predictive for cosmology. This is due to the fact that the particles
of the standard model and their interactions are well known, such that the flow
equations contain essentially no free parameters for the relevant values of the
cosmon field. Without a violation of quantum scale symmetry in the beyond
standard model sector the time evolution of dynamical dark energy is not
compatible with the precise cosmological constraints. In contrast, the simple
assumption of a slow logarithmic running in the beyond standard model sector
leads to models of growing neutrino quintessence. The rich and interesting
phenomenology of these models may well be compatible with observation. For
further progress in this direction one needs to identify which type of
fluctuations lead to a scale violation in the beyond standard model sector.

We have focused in this note on exact scaling solutions of quantum gravity
according to fundamental scale invariance~\cite{CWFSI}. It is remarkable that
this very restrictive setting may lead to a cosmology compatible with
observation. Relevant parameters for the flow away from the scaling solution
could induce a small number of additional parameters for the cosmon potential,
which will need to be investigated. For example, a shift of $u$ by a constant
could render $u$ positive for all values of $\chi$, resulting in positive $U_E$
for all values of the scalar field. This would still predict dynamical dark
energy, but change the characteristics of its evolution.

Our discussion of late dynamical dark energy has to be combined with an
investigation of the inflationary epoch. Inflation is also predicted by the
scaling solution of quantum gravity. The details of the inflationary epoch will
depend, however, on unknown particles with high masses, as for grand unified
models. For a given particle content the scaling solution of quantum gravity is
very predictive for inflation as well. The kinetial of the cosmon, as reflected
by $Z$, stops to evolve for large $\chi$ once the metric fluctuations and heavy
particles have decoupled. The value of $Z$ then depends on assumptions for
unknown particles, whose fluctuations determine its flow for small $\chi$. The
values of $Z$, $x_c$ and $x(y\subt{in})$ are, in principle, calculable from the
scaling solution of a model which covers all values of $\chi$. We do not want
here to stick to a definite setting for the particle physics in the ultraviolet
limit. Then $Z$, $x_c$ and $y(t\subt{in})$ are effectively free parameters, in
addition to the neutrino masses.

It is often believed that quantum gravity effects only affect very early
cosmology. In contrast, our findings reveal that quantum gravity can also be
very predictive for late cosmology. The severe constraints on the existence of
scaling solutions for all values of the cosmon field fix the scaling solution
for the cosmon potential. This potential is a key ingredient for the dynamics of
dark energy. It can no longer be assumed ad hoc for phenomenological purposes,
but rather becomes a calculable quantity. We hope that a large and fruitful
research field emerges from this combination of cosmology with quantum gravity.







\nocite{*}
\bibliography{refs}

\end{document}